\documentclass[onecollarge,natbib]{svjour2}
\bibpunct{[}{]}{;}{n}{}{,} 
\smartqed  
\usepackage{graphicx}
\journalname{Few-Body Systems}
\begin{document}

\title{Effective Field Theory and the Gamow Shell Model
}
\subtitle{The $\rm{^6He}$ Halo Nucleus}


\author{J. Rotureau        \and
        U. van Kolck  
}


\institute{J. Rotureau \at
              Department of Physics, University of Arizona, 
              Tucson, AZ 85721, USA
\\
\emph{Present address:} Fundamental Physics, Chalmers University of Technology, 
              412 96 G\"oteborg, Sweden\\ 
              \email{rotureau@chalmers.se}                       
           \and
           U. van Kolck \at
           Institut de Physique Nucl\'eaire,
           CNRS/Universit\'e Paris-Sud 11 (UMR8608), 
           F-91406 Orsay Cedex, France\\
           Department of Physics, University of Arizona, 
           Tucson, AZ 85721, USA}

\date{Received: date / Accepted: date}

\maketitle

\begin{abstract}
We combine Halo/Cluster Effective Field Theory (H/CEFT) and the 
Gamow Shell Model (GSM) to
describe the $0^+$ ground state of  $\rm{^6He}$ as a three-body halo system. 
We use two-body interactions for the 
neutron-alpha particle and two-neutron pairs
obtained from H/CEFT at leading order,
with parameters determined from scattering in the
p$_{3/2}$ and s$_0$ channels, respectively.
The three-body dynamics of the system is solved using the 
GSM formalism, where the continuum states are incorporated 
in the shell model valence space.
We find that in the absence of three-body forces
the system collapses, since
the binding energy of the ground state
diverges as cutoffs are increased. 
We show that addition at leading order of a three-body force 
with a single parameter
is sufficient for proper renormalization
and to fix the binding energy to its experimental value.
\end{abstract}

\section{Introduction}
\label{intro}
Nuclei located far away from the valley of $\beta$-stability 
display peculiar features that do not occur for well bound nuclei. 
The strong coupling to the continuum
manifests itself in
the existence of halo configurations, where some nucleons orbit far away
from a core of more tightly bound nucleons,
and of Borromean systems,
where removal of one nucleon is accompanied
by at least one more nucleon.
The neutron-rich Helium isotopes $^6\rm{He}$ and $^8\rm{He}$ 
offer two examples of such nuclei:
both are Borromean halos that
have no bound excited states. 
They also exhibit the ``binding-energy anomaly'', {\it i.e.}, 
higher one- and two-neutron
emission thresholds in $\rm{^8He}$ than in $\rm{^6He}$. 

Halo configurations are characterized by a 
large nuclear radius compared to the size 
of the tightly bound core or, equivalently, 
by a small nucleon separation energy
compared to the core binding energy.
The physics of halo nuclei is a perfect arena for the 
application of effective field theory (EFT). 
EFTs provide a powerful framework to exploit separation 
of scales in physical systems in order to
perform systematic, model-independent calculations. 
If, for example, the relative momentum $k$ of two particles 
is much smaller than the inverse range of their interaction,
$1/R$, using contact interactions
observables can be expanded in powers of $kR$
\cite{pauloreview}. 

The application of EFT to halo and cluster systems,
Halo/Cluster EFT (H/CEFT), was first exemplified
in low-energy neutron-alpha particle ($n\alpha$) 
scattering \cite{Bert}. 
Even though there is no bound state,
the $n\alpha$ $T$ matrix has a resonance pole at 
an energy $E_{5gs} \simeq 0.8$ MeV much smaller
than the $E_{4ex} \simeq 20$ MeV 
it takes to excite the alpha particle.
The physics in the p wave 
is closely related to halo dynamics,
because the ground state in $\rm{^5He}$
can in a first approximation be described as an  $n\alpha$ 
system in the p$_{3/2}$-wave configuration,
which has the qualitative
characteristics of a two-body halo nucleus.
H/CEFT captures these features at leading order (LO),
and provides at next-to-leading order 
a good description of $n\alpha$ 
scattering \cite{Bert}.
H/CEFT has also been successfully applied
to other dilute two-body systems such as
low-energy $\alpha\alpha$ \cite{HCEFT1}
and proton-$^7$Li \cite{HCEFT5} scattering,
radiative neutron capture on $^7$Li \cite{HCEFT3},
and the electric properties of $^{11}$Be \cite{HCEFT4}.

The natural follow up to Ref. \cite{Bert}
is to consider the next 
halo system within the Helium isotope family,
that is, 
$\rm{^6He}$. This is the aim of the present paper. 
Since the $\rm{^6He}$ ground state 
is bound by only $E_{6gs}\simeq -1.0$ MeV,
it is appropriate
to assume that it can be described as 
the three-body system $n+n+\alpha$.
The $n\alpha$ interaction is that studied in Ref. \cite{Bert},
while the $nn$ force is a contact interaction
determined by the low-energy $nn$ scattering parameters 
\cite{nnpionless,pauloreview}.

It is well understood that the physics of three-body systems 
can be much richer than the 
physics of its two-body subsystems. 
A famous example is the Efimov effect which occurs for 
non-relativistic particles with short-range interactions: 
if the s-wave scattering lengths of its subsystems 
are tuned to infinity, there can be an infinite sequence of 
three-body bound states that has an accumulation point
at the three-body threshold \cite{efimov}. 
A closely related phenomenon 
is the Thomas effect \cite{thom},
where a finite-range two-body potential that is 
only attractive enough to support a single two-body bound 
state can produce three-body bound states with arbitrarily large 
binding energies as the range goes to zero.
In EFT at a given order, the Thomas effect 
is a consequence of an inappropriate
omission of a three-body force.
As it has been demonstrated for s-wave two-body
interactions \cite{threebozos},
a three-body force is necessary and sufficient
for renormalization-group (RG) invariance at LO,
allowing three-body energies to be independent
of the ultraviolet regulator.
The parameter associated with this force then provides
a scale for the remaining discrete scale invariance, which
reflects itself in the Efimov spectrum.
The structure of $^{22}$C and other two-neutron halos
with s-wave neutron-core interactions 
was discussed using H/CEFT in Ref. \cite{HCEFT2}.

The situation with p-wave interactions is less clear.
There is debate over whether the Efimov effect 
can be realized in this case ---see Ref. \cite{luc}
and references therein.
The main issue to be addressed below is whether
a three-body force is needed at LO
so that the EFT description
of $\rm{^6He}$ is properly renormalized.

A microscopic description of weakly bound/unbound nuclei 
requires taking into account the interplay between 
bound states, scattering states, and resonances. In other
words, these systems have to be described as 
open quantum systems (OQSs),
in contradistinction with well-bound nuclei, 
which are nearly isolated from 
the environment of scattering states and decay channels
(``closed quantum systems'').  
A recent realization of the shell model for OQSs is the 
so-called Gamow Shell Model (GSM) \cite{GSM,GSM6He,GSM_6He,Hag}. 
The GSM is  based on the Berggren 
basis \cite{berg},
which consists of bound, resonant and  scattering 
single-particle wave functions generated by a 
finite-depth potential,
and provides the mathematical foundation for unifying 
bound and resonant states ---the poles of the $T$ matrix--- 
in the context of the Schr\"odinger equation. 
The GSM has been used \cite{GSM6He,GSM_6He,Hag} to study the properties
of the Helium isotope family using 
a phenomenological $n\alpha$ potential and a residual
two-neutron interaction adjusted to
few-body energies.

In this paper we use the formalism of the GSM to
solve the Schr\"odinger equation describing the dynamics 
of $\rm{^6He}$ with the contact interactions of H/CEFT,
including a possible three-body force.
We focus on the binding energy of the $0^+$ ground state.

The paper is organized as follows. 
In Sec. \ref{pot_sec} we review the potentials 
derived with EFT and used
for the study of $\rm{^6He}$. 
We introduce the 
GSM formalism in Sec. \ref{sec_GSM}, and
in Sec. \ref{sec_res} 
we show results for the ground-state energy of 
$^6 \rm{He}$. We shall see
that without a three-body force, the system is not properly 
renormalized at LO
\footnote{Our first results were presented
in Ref. \cite{jimmytalk}.
Similar results have been obtained 
independently by Ji, Elster, and Phillips \cite{jitalk}.}. 
We conclude and summarize in Sec. \ref{conc}.

\section{Two-Body Potentials}
\label{pot_sec}

Being weakly bound compared to the 
first excited state of the alpha particle, 
the $J^{\pi}=0^+$ ground state in $\rm{^6He}$ 
can be described as a three-body system $n+n+\alpha$. 
The neutrons in the halo interact with the alpha particle
via a two-body interaction $V_{n\alpha}$ and among themselves 
via a potential $V_{nn}$.
We denote the neutron (core) mass by $m_n$ ($M_c$)
and the neutron-core reduced mass by 
$\mu=m_n M_c/(m_n+M_c)$.

The potential between the $\alpha$ core and a neutron 
is constructed with EFT  as described in Ref. \cite{Bert}. 
The small relative momentum means that neutron and alpha
particle see each other, in a first approximation,
as elementary objects.
At LO there is only
one contribution, which is in the p$_{3/2}$ channel, 
and the ``dimeron'' potential projected onto this
channel can be written as
\begin{equation}
V_{n\alpha}(k',k,k_0)=
\frac{k' k}{A+B k_0^2},
\label{Vnaunreg}
\end{equation}
where $\vec{k}$ ($\vec{k'}$)  is the incoming (outgoing) 
relative momentum  
and $k_0=\sqrt{2 \mu E_{n\alpha}}$ in terms of
the total energy $E_{n\alpha}$ of the $n\alpha$ subsystem.
$A$ and $B$ are parameters. 
Since this interaction is singular, 
a regularization procedure is introduced in form of an
ultraviolet cutoff $\Lambda_{n\alpha}$.
The cutoff separates the short-distance physics, which is 
not included explicitly in the dynamics 
at low energies, and the long-distance physics, which is. 
This is here achieved by introducing 
a smooth regulator function 
\begin{equation}
F(x)=\exp(-x),
\end{equation}
whose role is to suppress the high-energy contributions 
of the potential. We thus replace the potential 
(\ref{Vnaunreg}) by
\begin{equation}
V_{n\alpha}(k',k,k_0;\Lambda_{n\alpha})=
\frac{k' k}{A(\Lambda_{n\alpha})+B(\Lambda_{n\alpha}) k_0^2} 
\, F\left(k'^2/\Lambda_{n\alpha}^2\right) 
\, F\left(k^2/\Lambda_{n\alpha}^2\right).
\label{pot_n_a}
\end{equation}
In order for observables to be 
RG invariant, 
{\it{i.e.}}, independent of the arbitrary cutoff,
the parameters 
$A(\Lambda_{n\alpha})$ and $B(\Lambda_{n\alpha})$ 
must depend on $\Lambda_{n\alpha}$.

More precisely, at LO,  $A(\Lambda_{n\alpha})$ and 
$B(\Lambda_{n\alpha})$ are fixed such that the phase
shifts at low energies obtained with the potential (\ref{pot_n_a}) 
reproduce the effective range expansion (ERE) in the
p$_{3/2}$ channel truncated
at the level of the effective ``range'':
\begin{equation} 
k^{3} \cot \delta_{n\alpha}(k) =
-\frac{1}{a_{n\alpha}}+\frac{r_{n\alpha}}{2} k^2,
\label{ERE_p}
\end{equation}
with the scattering volume 
$a_{n\alpha}=-62.951$ fm$^3$ and the effective momentum
$r_{n\alpha}=-0.8819$ fm$^{-1}$ \cite{param}.
The position $k_{res}$ of the p$_{3/2}$ resonance  
is obtained from 
\begin{eqnarray}
\cot \delta_{n\alpha} (k_{res}) =i,
\label{pole}
\end{eqnarray}
so at this order
$k_{res}=(0.174824 -0.031319 i)~\rm{fm^{-1}}$.
By solving the Lippman-Schwinger equation with the 
potential (\ref{pot_n_a}),  
one obtains 
\begin{eqnarray}
A(\Lambda_{n\alpha})&=& 
2 \mu\left [\frac{1}{a_{n\alpha}}
-\frac{\Lambda^3_{n\alpha}}{4\sqrt{2\pi}} \right ],
\\
B(\Lambda_{n\alpha})&=& 
-\mu\left[
r_{n\alpha}
+\frac{4}{a_{n\alpha}\Lambda^2_{n\alpha}}
+ \sqrt{\frac{2}{\pi}}\Lambda_{n\alpha}\right ].
\end{eqnarray}

The two neutrons in the halo have 
sufficiently low relative momentum
that meson exchange can be considered a short-range force. 
The neutron-neutron potential 
is thus taken from the pionless EFT
\cite{nnpionless,pauloreview}.
At LO, the potential is entirely in the $\rm{^1s_0}$ channel;
in momentum space it is simply a constant $C$.
As before, the potential requires regularization,
for which we continue to use the function $F(x)$,
but now in terms of the relative momentum between 
the two neutrons and of an $nn$ cutoff $\Lambda_{nn}$:
\begin{equation}
V_{nn}(k',k;\Lambda_{nn})=
C(\Lambda_{nn}) 
\, F\left(k'^2/\Lambda_{nn}^2\right)
\, F\left(k^2/\Lambda_{nn}^2\right).
\end{equation}
As previously, we fix the coupling constant 
$C(\Lambda_{nn})$ with the ERE for $nn$ scattering,
but now truncated at the level of the scattering length,
\begin{equation} 
k \cot \delta_{nn}(k) =
-\frac{1}{a_{nn}},
\label{ERE_s}
\end{equation}
with $a_{nn}=-18.7$ fm 
\cite{nnscatlength}. 
Again solving the Lippman-Schwinger equation,
\begin{equation}
C(\Lambda_{nn}) = 
\frac{1}{m_n} 
\left[ \frac{1}{a_{nn}} 
-\frac{\Lambda_{nn}}{\sqrt{2\pi}}\right] ^{-1}.
\end{equation}

Note that we do not modify the $nn$ potential in $^6$He
to account for the presence of the $\alpha$ core,
as frequently done \cite{GSM6He,GSM_6He}.
This modification is a three-body effect that
in EFT is represented by three-body forces,
which are present starting at some order,
since they are not forbidden by any symmetry.
We want to determine whether such a force is needed
at LO to renormalize the $n+n+\alpha$ system.

\section {Schr\"odinger Equation with the Gamow Shell Model}
\label{sec_GSM}

We now consider the solution of the Schr\"odinger equation 
for the $n+n+\alpha$ system with the Gamow Shell Model.
We use coordinates  
inspired by the 
Cluster Orbital Shell Model \cite{COSM,GSM_6He}:
$\vec{r}_{i}$ is the position of neutron $i=1,2$
relative to the $\alpha$ core,
and $\vec{p}_i$ the corresponding momentum.
The Hamiltonian of the $n+n+\alpha$ system
with the two-body interactions $V_{n\alpha}$ and $V_{nn}$ 
is written as 
\begin{equation}
H=\sum_{i=1}^{2}
\left[\frac{\vec{p}_i^{\, 2}}{2 \mu}
+V_{n\alpha}(k_{0i};\Lambda_{n\alpha})\right]
+V_{nn}(\Lambda_{nn})+\frac{\vec{p}_1\cdot  \vec{p}_2}{M_c}. 
\label{ham_3b}
\end{equation}
This Hamiltonian is translationally invariant, 
the recoil term 
$\vec{p}_1 \cdot  \vec{p}_2/M_c$   
stemming from the choice of coordinates.

We work within the framework of the Gamow Shell Model 
formalism \cite{GSM,GSM6He,GSM_6He,Hag} 
to solve the dynamics generated by the Hamiltonian (\ref{ham_3b}).
The three-body equation is solved using a single-particle (sp)
basis.
The set of sp states that define
the one-body valence space is taken as the set of eigenstates 
of the LO potential $V_{n\alpha}(k_0;\Lambda_{n\alpha})$.
They are solutions of the one-body Schr\"odinger equation
\begin{equation}
H^{sp}|\Psi\rangle=
\left[\frac{\vec{p}^{\, 2}}{2 \mu}+V_{n\alpha}(k_0;\Lambda_{n\alpha})\right]
|\Psi\rangle
=E_{n\alpha}|\Psi\rangle .
\label{ham_1b}
\end{equation}
By inserting the completeness relation projected on the p$_{3/2}$ partial wave,
\begin{equation}
\int_{\cal C} dk \, k^2 \, |k\rangle \langle k|
=1, 
\label{comp}
\end{equation}
along a contour $\cal{C}$ in the fourth quadrant of the 
complex momentum plane,
Eq. (\ref{ham_1b}) can be written as an
equation for the momentum-space
wave function $\Psi(k)=\langle k|\Psi\rangle$,
\begin{equation}
\int_{\cal{C}}dk' k'^2 \, \langle k|
\left[\frac{p^2}{2 \mu}+V_{n\alpha}(k_0;\Lambda_{n\alpha})\right]
|k'\rangle \, \Psi (k')
=E_{n\alpha} \, \Psi (k).
\label{ham_1b_bis}
\end{equation}

In this paper 
the contour ${\cal C}$ is chosen to be made out of three
straight-line segments ${\cal C}_{1,2,3}$,
${\cal C}={\cal C}_{1} +{\cal C}_{2}+{\cal C}_{3}$.
Segment ${\cal C}_{1}$ extends
from $k_0=0$ to  $k_1=k_{1r} +i k_{1i}$, 
segment ${\cal C}_{2}$ from $k_1$
to $k_2=k_{2r}$,
and segment ${\cal C}_{3}$ from 
$k_2$ to $k_3=k_{max}$, where 
$k_{max}\ge k_{2r}\ge k_{1r}\ge 0\ge k_{1i}> -k_{1r}$
are real numbers.
Since the sp set must be finite, 
the contour integral along $\cal{C}$ is performed
up to a cutoff $k_{max}$ and discretized with a 
quadrature method. 
In this case, $k_{max}$ must be chosen large enough
such that all low-energy physics below $\Lambda_{n\alpha}$ 
is taken into account. Here,
we typically chose $k_{max}\sim 3\Lambda_{n\alpha}$.  
Had we chosen a sharp regulator for $F(x)$, 
$k_{max}$ would have been such that $k_{max}=\Lambda_{n\alpha}$
since in that case $V_{n\alpha}(k_0,\Lambda_{n\alpha})$ 
would have vanished for momentum above $\Lambda_{n\alpha}$.
In practice,
the contour $\cal{C}$ is discretized using a Gauss-Legendre 
quadrature using $N_i$ points for the segment ${\cal C}_{i}$,
for a total number $N_{sh}=N_1+N_2+N_3$ of discretization points.

\begin{figure}[t]
\centering
\includegraphics[scale=0.4,angle=-90,clip=true]{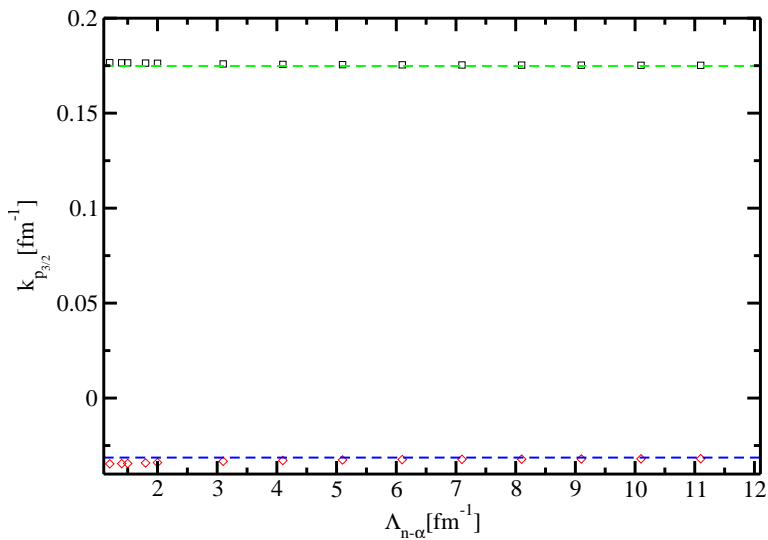}
\caption{Position of the lowest $\rm{p_{3/2}}$ $n\alpha$ resonance as a 
function of the cutoff $\Lambda_{n\alpha}$. 
Points at the top (bottom) are the results 
for the real (imaginary) momentum of the resonance coming
from the
numerical solution of the Schr\"odinger equation
described in the text.
The dashed lines are the  corresponding momenta
obtained from the effective range expansion with
empirical values for the scattering volume
and effective 
momentum.
}
\label{res_fig}
\end{figure}

If $k_{1i}=0$, the contour is along the real axis,
and the 
solutions of Eq. (\ref{ham_1b_bis}) consist of bound states 
and scattering states. 
If $k_{1i}\ne0$, solutions 
consist instead of bound states, resonant states located above 
the contour, and complex-scattering scattering states along
the contour \cite{berg}. 
In order to include the
p$_{3/2}$ resonance, we take
$k_{1r}=0.18$ fm$^{-1}$, 
$k_{1i}=-0.08$ fm$^{-1}$, 
and
$k_{2r}=0.5$ fm$^{-1}$.
We show in Fig. \ref{res_fig} the position of the p$_{3/2}$ 
resonance 
as a function of 
$\Lambda_{n\alpha}$. 
For each value of 
$\Lambda_{n\alpha}$ the Schr\"odinger 
equation (\ref{ham_1b_bis}) is solved with 
the LO potential along the 
complex contour $\cal{C}$
described above.
Results for the energy of the resonance are independent 
of the choice of the contour as long as it goes below the 
resonance and as long as the discretization is precise enough. 
For instance, for $\Lambda_{n\alpha}=3.1$ fm$^{-1}$ 
we use $k_{max}=10$ fm$^{-1}$, $N_1=10$, $N_2=10$, and $N_3=15$.
For comparison, 
$k_{res}$ obtained directly from the ERE
is also shown.
As it can been seen from the figure, 
as the cutoff  $\Lambda_{n\alpha}$ increases
the position of the resonance quickly converges to $k_{res}$.

Since the $\rm{^6He}$ ground state is bound, it is equivalent 
to use for the sp states 
either a set of shells located along the real continuum axis, 
or a complex-continuum set of states along a 
complex contour $\cal{C}$ along with the $p_{3/2}$ resonance. 
In the three-body calculations presented below we have used shells 
located along the real energy axis, that is,
we have taken $k_{1i}=0$.
For instance, for 
$\Lambda_{n\alpha}=7.1$ fm$^{-1}$,
we use $k_{1r}=5.0$ fm$^{-1}$, 
$k_{2r}=12.0$ fm$^{-1}$, and
$k_{max}=21.0$ fm$^{-1}$, 
with $N_1=N_2=N_3=30$.
We could certainly decrease the number of shells to reach 
the same precision in the value of the $\rm{^6He}$ ground state.
We have not studied in detail what would be the 
smallest admissible number of points.
Indeed, for a system made of three particles
this is not of a great importance,
since the diagonalization of the non-symmetric Hamiltonian matrix 
can be performed rather fast.

{}From the set of eigenstates of the Schr\"odinger equation 
(\ref{ham_1b}) a sp basis is generated.
For an energy-independent potential,
a resonant state $|\Psi_{res}\rangle$ above,
and scattering states along, ${\cal{C}}$ 
satisfy the usual Berggren relation \cite{berg},
\begin{equation}
|\Psi_{res}\rangle \langle \widetilde{\Psi}_{res}|
+\int_{\cal{C}} dk \, k^2 \, |\Psi(k)\rangle 
\langle \widetilde{\Psi(k)}|
=1,
\label{bergrel}
\end{equation}
where 
the bra $\langle \widetilde{\Psi}|$ conjugate to the ket
$|\Psi\rangle$ is such that
$\langle \widetilde{\Psi}|r\rangle=
\langle r|\Psi\rangle$.
However,
the potential $V_{n\alpha}(k_0; \Lambda_{n\alpha})$ 
being energy-dependent, 
the eigenstates of 
Eq. (\ref{ham_1b_bis}) are not orthogonal and 
Eq. (\ref{bergrel}) does {\it not} hold. 
One then has to consider an extra step to generate a basis. 
This is achieved, after having discretized the contour 
$\cal{C}$, by solving the matrix equation
\begin{equation}
\sum_{i=1}^{N_{sh}}{|\Psi_{i}}\rangle \langle  \Psi^{boc}_{i}|
=1, 
\label{sp1_1}
\end{equation}
where $|\Psi_{i} \rangle$ is one of the $N_{sh}$ 
discrete sp eigenstates 
of the potential, Eq. (\ref{ham_1b}), and 
$\langle  \Psi^{boc}_{i} |$ its bi-orthogonal complement. 
By construction,
\begin{equation}
\langle \Psi^{boc}_{i} |\Psi_{j} \rangle=\delta_{ij}.
\end{equation}

A complication is that for sufficiently large values of the cutoff, 
$\Lambda_{n\alpha} \ge \Lambda_b \simeq 1.8$ fm$^{-1}$, the  
potential $V_{n\alpha}(k_0;\Lambda_{n\alpha})$ supports 
a bound state $|\Psi_{b}\rangle$. 
At $\Lambda_{n\alpha} = 1.8$ fm$^{-1}$ the energy of this 
bound state is $E_b=-20.941$ MeV, that is, outside the 
range of validity of our EFT approach. 
As a consequence, we do not want to include it in the valence space. 
{}From the practical point of view, we first tried to construct
the bi-orthogonal basis
by including the bound state in the bi-orthogonalization procedure,
and then omitting it when constructing the many-body basis to solve 
the three-body problem. 
This procedure turned out to give rather peculiar results in the sense that  
the value obtained for the $\rm{^6He}$ ground state 
displayed a discontinuous behavior as  
$\Lambda_{n\alpha}$ varied from values below to above $\Lambda_{b}$. 
We suspect that despite the fact that the bound state is not included
in the basis, it has an indirect effect,
for it is present during the phase
of construction of the sp basis according to Eq. (\ref{sp1_1}). 
One has then to figure out
another way to generate the sp shells when 
a deep bound state is present.

For $\Lambda_{n\alpha} \ge \Lambda_b$, 
we generate the sp basis by converting the energy dependence 
of $V_{n\alpha}(k_0;\Lambda_{n\alpha})$ into 
momentum dependence by introducing 
an energy-independent potential 
$V'_{n\alpha}(\Lambda_{n\alpha})$ that reproduces the  
half-on-shell $T$ matrix \cite{Bogner},
\begin{equation}
\langle k'|V'_{n\alpha}(\Lambda_{n\alpha})|\Psi
\rangle
=\langle k'|V_{n\alpha}(k_{0};\Lambda_{n\alpha})|\Psi
\rangle, 
\label{pot_k}
\end{equation}
where $k_{0}$ is obtained from the energy of
the $H^{sp}$ eigenstate $|\Psi\rangle $.
For each discretized value $k'_i$ along the contour $\cal{C}$,
we solve Eq. (\ref{pot_k}) in order
to generate $V'_{n\alpha}(k'_i,k_j;\Lambda_{n\alpha})$ 
{\it{without}} considering the bound state $|\Psi_{b}\rangle$. 
For each $k'_{i}$ we have $N_{sh}$ unknowns, 
$V'_{n\alpha}(k'_i,k_j;\Lambda_{n\alpha})$
with $j=1,\ldots,N_{sh}$, and $N_{sh}-1$ equations,
\begin{equation}
\langle k'_{i}|V'_{n\alpha}(\Lambda_{n\alpha})|\Psi_{j}\rangle
=\langle k'_{i}|V_{n\alpha}(k_{0j};\Lambda_{n\alpha})|\Psi_{j}\rangle,
\quad j=1,\ldots,N_{sh}, 
\quad j \ne b.
\end{equation}
In order to solve this linear system we impose the condition 
\begin{equation}
\langle k'_{i}|V'_{n\alpha}(\Lambda_{n\alpha})|k_{N_{sh}}\rangle=0,
\end{equation}
with $|k_{N_{sh}}\rangle$ being the state with 
the largest momentum on the contour $\cal{C}$. 
This leads to a small error, since at such high momentum, 
$k_{N_{sh}}\sim {k_{max}} \sim 3 \Lambda_{n\alpha}$,
the influence of the potential is negligible.
Moreover this error can be made arbitrarily
small by increasing $k_{max}$.
The potential 
$V'_{n\alpha}(\Lambda_{n\alpha})$ is non-Hermitian and has  
right eigenvectors $|\Psi_{i}\rangle$  and  left eigenvectors  
$\langle \Psi^{left}_{i}|$. 
The right eigenvectors are by construction the eigenvectors of the original 
energy-dependent potential $V_{n\alpha}(k_0;\Lambda_{n\alpha})$,
and we now have the following completeness relation:
\begin{equation} 
\sum{|\Psi_{i}}\rangle \langle \Psi^{left}_{i}|=1. 
\label{sp1}
\end{equation}

For $\Lambda_{n-\alpha} < \Lambda_b$, the two previous procedures 
for constructing the sp basis are completely identical,
the left eigenvectors $\langle \Psi^{left}_{i}|$ of 
$V'_{n\alpha}(\Lambda_{n\alpha})$ obtained 
with the second method being equal to the bi-orthogonal 
complement states $\langle  \Psi^{boc}_{i}|$ 
obtained in the first method by 
solving Eq. (\ref{sp1_1}). 

{}From the sp basis, we construct the antisymmetrized 
three-body basis states coupled to good total angular momentum $J$,
$|(\Psi_{i},\Psi_{j})^{J}_{i \leq j}\rangle$, 
which are eigenstates 
of the
Hamiltonian $H^{sp}_{1}+H^{sp}_{2}$
with eigenvalues $E_i+E_j$:
\begin{equation} 
\left(H^{sp}_{1}+H^{sp}_{2}\right)|(\Psi_{i},\Psi_{j})^{J}_{i \leq j}\rangle=
\left(E_i+E_j\right)|(\Psi_{i},\Psi_{j})^{J}_{i \leq j}\rangle.
\end{equation}
The corresponding bi-orthogonal complement is 
$\langle({\Psi^{left}_{i}},{\Psi^{left}_{j}})^{J}_{i \leq j}|$.

The interaction 
$V_{nn}(\Lambda_{nn})$ is defined in terms of relative 
coordinates between the two neutrons. 
Since our Hamiltonian was 
written in terms of $n\alpha$ coordinates,  a transformation
is necessary to express the matrix elements $V_{nn}(\Lambda_{nn})$ 
in the shell model basis $|(\Psi_{i},\Psi_{j})^{J}_{i \leq j}\rangle$. 
For this purpose we use an expansion on a set of 
harmonic-oscillator (HO)
wave functions, as in Ref. \cite{Hag}. 
That is, we project $V_{nn}(\Lambda_{nn})$ 
on a HO set $| ab\rangle$,
where $a$ and $b$ label sp states of HOs
in the $n\alpha$ coordinate,
and consider 
the $nn$ interaction
\begin{equation}
V_{nn}^{osc}(\Lambda_{nn})=
\sum_{a<b}\sum_{c<d} 
|ab\rangle \langle ab|V_{nn}(\Lambda_{nn})|cd\rangle \langle cd|,
\label{Vnnosc}
\end{equation}
where the restriction in the sum is due to the antisymmetry
of the two-neutron state.
Using Moshinsky transformations \cite{mosh}, one can easily calculate 
$\langle ab|V|cd\rangle$.
Results for the three-body energy are independent of the 
values of the HO frequency,
as long as enough HO states are included in the expansion.

\section{Results for the $\rm{^6He}$ Ground State}
\label{sec_res}

The ground state of $\rm{^6He}$ is coupled to $J^{\pi}=0^+$ 
and the three-body basis states are constructed from the 
sp states of the  $V_{n\alpha}(k_0;\Lambda_{n\alpha})$ potential 
as described in the previous section. 
At LO, only p$_{3/2}$ shells are included in the valence
space and, as a consequence, all matrix elements of the 
recoil term in Eq. (\ref{ham_3b}) vanish. 
For each value of $\Lambda_{n\alpha}$ the coupling constants 
$A(\Lambda_{n\alpha})$ and $B(\Lambda_{n\alpha})$ are fixed 
such that the ERE in the p$_{3/2}$ channel truncated at the 
level of the effective ``range'' is reproduced. 
Similarly, $C_0(\Lambda_{nn})$  is fixed
such that the $^1$s$_0$ $nn$ scattering length
is reproduced.

\begin{figure}[t]
\centering
\includegraphics[scale=0.4,angle=-90,clip=true]{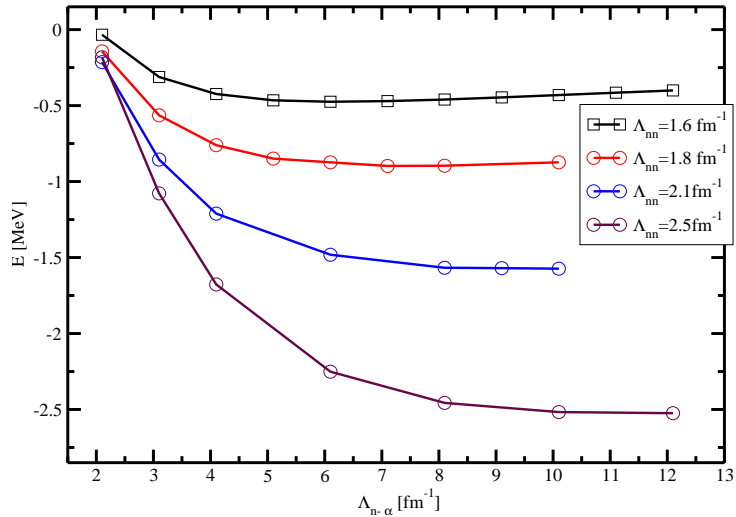}
\caption{Ground-state energy of $\rm{^6He}$ 
from the LO two-body potentials 
$V_{n\alpha}(k_0;\Lambda_{n\alpha})$
and $V_{nn}(\Lambda_{nn})$. 
For each value of $\Lambda_{nn}$ 
the cutoff $\Lambda_{n\alpha}$ is varied.}
\label{energy_dif_cut}
\end{figure}

Figure \ref{energy_dif_cut} shows the energy $E_{nn\alpha}$
of the ground state in $\rm{^6He}$  for different values of 
$\Lambda_{n\alpha}$ and $\Lambda_{nn}$. 
For each value of $\Lambda_{nn}$  
the cutoff $\Lambda_{n\alpha}$ is increased. 
One can see 
that the energy initially quickly decreases,
then slowly rises.
For $\Lambda_{nn}=1.6$ fm$^{-1}$, for example,
$E_{nn\alpha}$
goes from $-0.034$ MeV for $\Lambda_{n\alpha}=2.1$ fm$^{-1}$ to 
$-0.475$ MeV for $\Lambda_{n\alpha}=6.1$ fm$^{-1}$, 
then to 
$-0.400$ MeV for $\Lambda_{n\alpha}=12.1$ fm$^{-1}$.
As $\Lambda_{nn}$ increases,
the initial decrease 
becomes steeper, and the increase is postponed to
higher values of $\Lambda_{n\alpha}$.
For instance, at $\Lambda_{nn}=2.5$ fm$^{-1}$,
the energy goes from $-0.182$ MeV to $-2.251$ MeV 
to $-2.524$ MeV
in the same range of $\Lambda_{n\alpha}$ values. 

This behavior can be understood 
from the qualitative renormalization features of the system.
As $\Lambda_{n\alpha}$ increases, the phase
space of the three-body system increases,
the attractive $nn$ interaction is better resolved,
and the binding energy increases.
This is consistent with
the pattern observed in Ref. \cite{us_trap} for the energy 
of a three-fermion system interacting via a two-body force 
constructed with EFT at LO. In that case, for a fixed cutoff
of the two-body interaction, the total energy of the system 
decreases as the size of the model space increased. 
As $\Lambda_{nn}$ increases, presumably
more correlations are cut off for too small a value of 
$\Lambda_{n\alpha}$,
generating the faster decrease.
However, there is also a residual
dependence on $\Lambda_{n\alpha}$ 
from $V_{n\alpha}(k_0;\Lambda_{n\alpha})$.
Even though the potential 
has been properly renormalized, that is, 
the coupling constants $A(\Lambda_{n\alpha})$ and 
$B(\Lambda_{n\alpha})$ 
have been fixed so that the truncated ERE 
is reproduced, there still is a dependence for finite 
values of the cutoff, as seen in Fig. \ref{res_fig}.
The energy of the p$_{3/2}$ resonance goes from 
$k=0.7714 -0.2947i$ MeV to 
$k=0.7696 -0.2896i$ MeV when 
$\Lambda_{n\alpha}$ goes from 
$6.1$ fm$^{-1}$ to $12.1$ fm$^{-1}$. 
This means that, as  $\Lambda_{n\alpha}$ is varied
within this range, there is a variation
$\simeq 0.005$ MeV, or about 7\%,
in the norm of the energies of 
the p$_{3/2}$ resonance,
which is consistent with a 
variation of about 15\% in the three-body energy 
in the same range
---for example a variation of 
$\simeq 0.075$ MeV for $\Lambda_{nn}=1.6$ fm$^{-1}$.

\begin{figure}[t]
\begin{center}
\includegraphics[scale=0.4,angle=-90,clip=true]{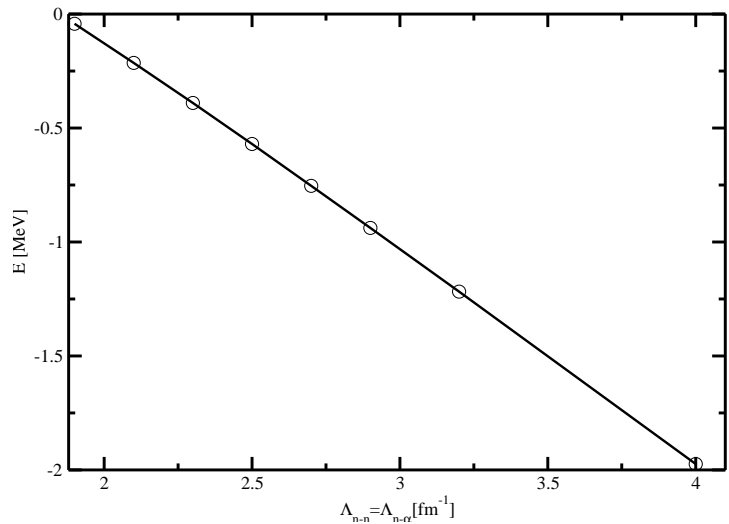}
\end{center} 
\caption{Ground-state energy of $\rm{^6He}$ 
from the LO two-body potentials 
$V_{n\alpha}(k_0;\Lambda_{n\alpha})$
and $V_{nn}(\Lambda_{nn})$,
for $\Lambda_{nn}=\Lambda_{n\alpha}$.}
\label{energy_same_cut}
\end{figure}

One can clearly see from Fig. \ref{energy_dif_cut}
that as the cutoffs $\Lambda_{nn}$ and $\Lambda_{n\alpha}$
are increased, the energy decreases
without reaching a stabilized value. 
To stress this fact,
in 
Fig. \ref{energy_same_cut}
we plot the $\rm{^6He}$ ground-state energy
as function of $\Lambda_{nn}=\Lambda_{n\alpha}$.
We have checked that the results are
similar if other relations are assumed
between  $\Lambda_{nn}$ and $\Lambda_{n\alpha}$,
for example, if we take the
minimum energy for each $\Lambda_{nn}$,
which is equivalent to choosing
$\Lambda_{n\alpha}$ large enough so that all correlations 
of the $nn$ interaction have been resolved by the 
three-body system.

The nearly linear dive of the ground state seen in
Fig. \ref{energy_same_cut}
is reminiscent of the behavior
observed with LO two-body forces in EFTs for systems of
three bosons or three-or-more-component fermions
\cite{threebozos}.
There, the dive is even faster, more like quadratic
in the cutoff, stemming from the 
strong s-wave interactions among the three pairs.
In either case, what we see is a collapse
of the ground state under short-range two-body interactions
similar to the one first observed by Thomas \cite{thom}.
It is 
an indication that the three-body problem has not been 
properly renormalized with only two-body interactions 
\cite{threebozos}. 
The cutoff dependence is not decreasing
as the cutoffs increase, as one would expected from
residual cutoff dependence in a 
renormalized system that has been truncated correctly,
but is instead increasing with positive
powers of the cutoffs.

The solution to this problem has to be found
outside the two-body subsystems, which are perfectly well 
defined and well described by the EFT.
We thus add a three-body force to renormalize 
the three-body problem. 
An s-wave three-body force 
does not have any impact at LO on the structure of 
$\rm{^6He}$ since the $n\alpha$ subsystems 
are in a relative p$_{3/2}$ wave. 
The lowest-derivative three-body force
that does not vanish in the channel of interest
can be written, in the coordinates we are using, as
\begin{equation}
V_{nn\alpha}(k'_1,k'_2,k_1,k_2)=
D(\Lambda_{nn\alpha})
\, k'_1 k_1
k'_2 k_2 
\, F\left(k'^2_1/\Lambda_{nn\alpha}^2\right) 
F\left(k'^2_2/\Lambda_{nn\alpha}^2\right)
F\left(k_1^2/\Lambda_{nn\alpha}^2\right) 
F\left(k_2^2/\Lambda_{nn\alpha}^2\right),
\label{3b_force}
\end{equation}
with $\vec{k'}_i$ ($\vec{k}_i$) the outgoing (incoming)
momentum for the $i$th $n\alpha$ subsystem.
Here $\Lambda_{nn\alpha}$ is a three-body cutoff 
and $D$ is a low-energy coupling constant
with dimensions of mass$^{-9}$, whose
dependence on $\Lambda_{nn\alpha}$ is adjusted so 
that three-body observables be (nearly)
cutoff independent.

\begin{figure}[t]
\begin{center}
\includegraphics[scale=0.4,angle=-90,clip=true]{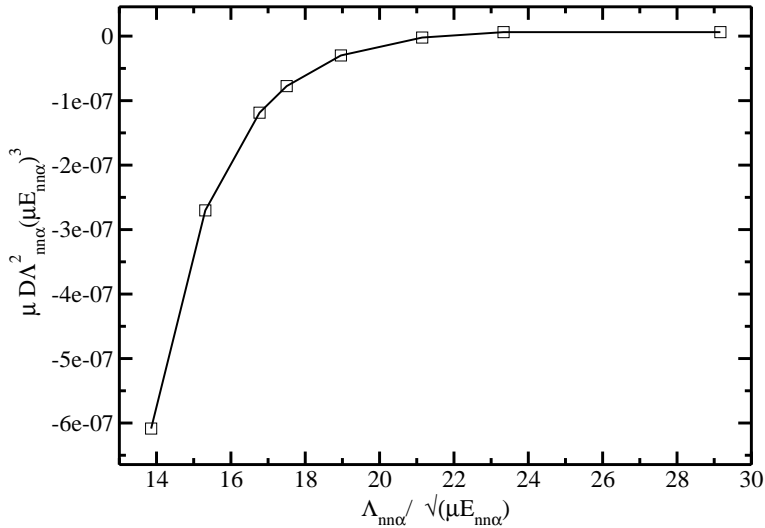}
\end{center} 
\caption{Dimensionless three-body coupling constant 
$\mu^4 E_{nn\alpha}^3 \Lambda_{nn\alpha}^2 D$
as function of $\Lambda_{nn\alpha}/\sqrt{\mu E_{nn\alpha}}$, 
when the
$\rm{^6He}$ ground-state energy $E_{nn\alpha}$
is fixed at its experimental value, for
$\Lambda_{nn\alpha}=\Lambda_{nn}=\Lambda_{n\alpha}$.
}
\label{3b_fig}
\end{figure}

Here for simplicity 
we take $\Lambda_{nn\alpha}=\Lambda_{nn}=\Lambda_{n\alpha}$.
We find that we can then keep the $\rm{^6He}$ ground-state
energy $E_{nn\alpha}$ constant as the cutoff is varied.
We show in Fig. \ref{3b_fig} the resulting running of the 
coupling constant $D(\Lambda_{nn\alpha})$ 
when the $\rm{^6He}$ ground-state energy
is fixed to its experimental value 
$E_{6gs}=-0.98$ MeV \cite{6Hegsenergy}.
At low cutoffs, $D$ is negative.
{}From Fig. \ref{energy_same_cut} we see
that at a cutoff
$\Lambda_{0}\simeq 2.9$ fm$^{-1}$
the energy calculated with
only two-body forces agrees with the experimental
value, so $D(\Lambda_{0})=0$.
Above $\Lambda_{0}$,
$D(\Lambda_{nn\alpha})\Lambda_{nn\alpha}^2$
is positive and approximately constant in the region 
of cutoffs we could probe.
We cannot, however, exclude a limit-cycle-like
behavior at higher cutoffs,
as observed in Ref. \cite{threebozos}.

Again like for
three bosons or three-or-more-component fermions
\cite{threebozos},
RG invariance requires the three-body force
to appear at LO.
Naturalness together with naive dimensional analysis
suggests that $D$ would scale as $M^{-9}$,
with $M$ a large mass scale such as the 
alpha-particle binding momentum or the pion mass.
If that were the case after renormalization, 
the three-body force
(\ref{3b_force}) would be a very high-order effect.
Instead here,  
as for the three-nucleon system \cite{threebozos},
a certain amount of fine-tuning is present:
the low-energy
scale responsible for the existence of
the shallow two-nucleon $^1$s$_0$ virtual
bound state and the shallow $^5$He p$_{3/2}$ resonance
must
appear in the renormalized three-body force as well.
The infrared enhancement of the LO two-body interactions
dominates the running of the LO three-body force,
making its effects much larger than the naturalness expectation.
While in the pure s-wave case the enhancement is 
proportional to the square of the large scattering length \cite{threebozos},
here it must be roughly the square of the large 
scattering volume.

With the three-body force so determined,
we have looked for other $0^+$ bound states and
found none within the cutoff range we investigated. 
This is perhaps not surprising.
It has been argued that the Efimov effect \cite{efimov}
is present
if both the
scattering volume and the effective momentum in a
system with pairwise p-wave interactions
are large,
although there is debate about whether this can be realized
\cite{luc}.
Since $r_{n\alpha}$ is not particularly
large, we would not expect here an Efimov tower of shallow
three-body states anyway.

\section{Conclusions and Outlook}
\label{conc}

In this paper we have described for the first 
time the ground state of 
$\rm{^6He}$ using interactions derived from 
Halo/Cluster Effective Field Theory, 
where the alpha-particle core is treated as 
an explicit field \cite{Bert}. 
The two-body $n\alpha$ and $nn$ interactions
are of the contact type, with parameters
determined from two-body scattering data. 
The three-body dynamics of the system was solved using the 
formalism of the Gamow Shell Model \cite{GSM}, 
where the set of single-particle states 
(resonant and continuum) is given by the $n\alpha$ potential.
We had to adapt the formalism to accommodate
the energy dependence of the LO $n\alpha$ EFT potential.
This is also the first time the GSM has been applied to 
the solution of EFT.
 
We have seen that, at leading order, two-body forces are not sufficient 
to properly renormalize the three-body system,
even though they provide a systematic
expansion for two-body scattering \cite{Bert}. Indeed, as 
the cutoffs are increased the energy of the three-body ground state 
does not stabilize and would collapse for an arbitrarily 
large cutoff.
We have shown that the addition of a single
three-body counterterm is enough to 
achieve renormalization-group invariance
\footnote{It is our understanding that the same conclusion
was reached by Ji, Elster and Phillips \cite{jitalk,danpers}.},
as for systems with s-wave interactions \cite{threebozos}.
We have obtained the RG running
of the coupling constant by demanding
that the binding energy be fixed at its experimental value.

Our work paves the way for more comprehensive
studies of halo nuclei with H/CEFT.
For the future, we plan to carry out a more extensive
investigation of $\rm{^6He}$, including 
higher-order corrections
and calculation of other observables
(such as the ground-state radius and the 
first excited-state energy). 
At the cost of more computational resources,
other members of the $\rm{He}$ isotope family could be 
investigated as well, along the lines of 
Refs. \cite{GSM6He,GSM_6He}.
More generally, we hope that 
the combination of EFT and GSM 
will prove to be a valuable tool in the study of
other three-body resonant states,
such as the Hoyle state in $\rm{^{12}C}$.

\begin{acknowledgements}
We thank Scott Bogner and Georgios Papadimitriou for useful discussions,
and Daniel Phillips for interesting comments on the manuscript.
This research was supported in part 
by the European Research Council (ERC StG 240603) under the FP7 (JR),
the US NSF under grant PHY-0854912 (JR),
and the US DOE under grant DE-FG02-04ER41338 (JR and UvK).
\end{acknowledgements}



\end{document}